\documentclass[final,english]{bullsrsl}

\usepackage[latin1]{inputenc}

\usepackage[T1]{fontenc}

\usepackage[numbers]{natbib}

\usepackage{graphicx}
\usepackage{rotating}

\begin{document}
\title{Interacting supernovae from wide mass-transferring binaries}





\author[]{Andrea}{Ercolino}

\affiliation{Argelander Institut f\"ur Astronomie, Auf dem H\"ugel 71, DE-53121 Bonn, Germany}


\correspondance{aercolino@astro.uni-bonn.de}


\maketitle

\begin{abstract}
The light curves and spectra of many Type I and Type II supernovae (SNe) are heavily influenced by the interaction of the SN ejecta with circumstellar material (CSM) surrounding the progenitor star. The observed diversity shows that many progenitors have undergone some level of stripping and polluted their CSM shortly before the explosion. The presence of a binary companion and the mass transfer that can ensue offers a mechanism that can give rise to this diversity.
We present a set of detailed massive evolutionary models in which the donor star, a Red Supergiant (RSG) is in a wide orbit around a main-sequence companion, and undergoes stable or unstable mass transfer in the later stages of evolution, up to the moment of core collapse.  We also discuss some significant physics of these systems that may impact our results, from the presence of pulsations and extended atmospheres in RSGs, to the initial eccentricity of the
orbit. 
The resulting SN types range from Type IIP to H-deficient IIb and H-free Ib. 
In models undergoing stable mass transfer, the material lost during this process is expected to form a dense CSM surrounding the system by the time of core collapse and give rise to significant interaction effects in the SN light curve and spectra.
In the systems with unstable mass transfer mass transfer, the SN may occur during common-envelope evolution. In this case, the progenitor may show significant variability in the last few thousand years before core collapse, and the following SN will likely exhibit strong interaction effects.
\end{abstract}


\section{Introduction}
Supernova (SN) explosions are the end of the life of massive stars, and the observation of these phenomena has helped assess our understanding of the last evolutionary phases of massive stars. In the last few decades, the surveys currently in place (ASAS-SN \cite{ASASN}, ATLAS \cite{ATLAS}, Pan-STARRS \cite{PanSTARRS}, PTF \cite{PTF},  ZTF \cite{ZTF}) increased the number of observations of so-called `interacting SNe', that is SNe which interact with the surrounding circumstellar material (CSM).

The main indicator of SN-CSM interaction is the presence of narrow-line emission spectra which indicate a relatively slow-moving material compared to the fast-expanding ejecta. The light-curve may also be affected, as the collision between the ejecta and the CSM provides an additional energy source (e.g. \cite{Fransson_2010jl, Smith2008_2006gy}). The timescales for these interaction features are varied, as in some cases they are present only at early times \cite{Yaron_Flash}, late times \cite{Margutti2017_2014C} or even throughout the entire evolution \cite{Leonard_1998S}. The mechanisms to produce such a variety of CSM structures are still unclear.

To explain the presence of this CSM, many works invoke a dramatic phase of mass-loss before the SN explosion of the progenitor star, induced by instabilities (like LBV eruptions \cite{SmithArnett2014_Hydroinstab_Turb_preSN}) or wave-driven envelope excitations (e.g., \cite{Quataert_Shiode_wavedriven_winds,Woosley_Heger_2015_SiFlash, Fuller17_waveheating_RSG}).
Mass transfer (or Roche lobe overflow, RLOF) during the later stages of evolution with a companion star can also produce the CSM, so long as most of the mass fails to be accreted. This is a promising channel, as most massive stars are found in binaries \cite{Sana_massive_stars_binaries}.

In this paper we report the results shown in \cite{Ercolino_widebinary_RSG} where we discussed late-time mass transfer from a Red Supergiant (RSG) primary to a main-sequence (MS) secondary star in the context of interacting H-rich SNe.

\section{Results}
A series of binary evolutionary models were run with the code \texttt{MESA} \citep{MESA_I, MESA_II, MESA_III, MESA_IV, MESA_V}, until core-collapse (CC) of the initially more massive star, with a mass of $12.6M_\odot$. The models comprised binaries with companion masses between $12.0$ and $1.3M_\odot$ and initial orbital periods between $562$ and $2818\,\mathrm{d}$. The main physics and parameters are mentioned in \citep{Ercolino_widebinary_RSG}. {  The models are constructed such that they experience Case C RLOF in the last $\sim20\,\mathrm{kyr}$ before CC. The companion typically accretes only $\sim0.1M_\odot$, while the rest of the material transferred is assumed to be lost from the system thus forming the CSM.} 

\begin{figure}
    \centering
    \includegraphics[width=1\linewidth]{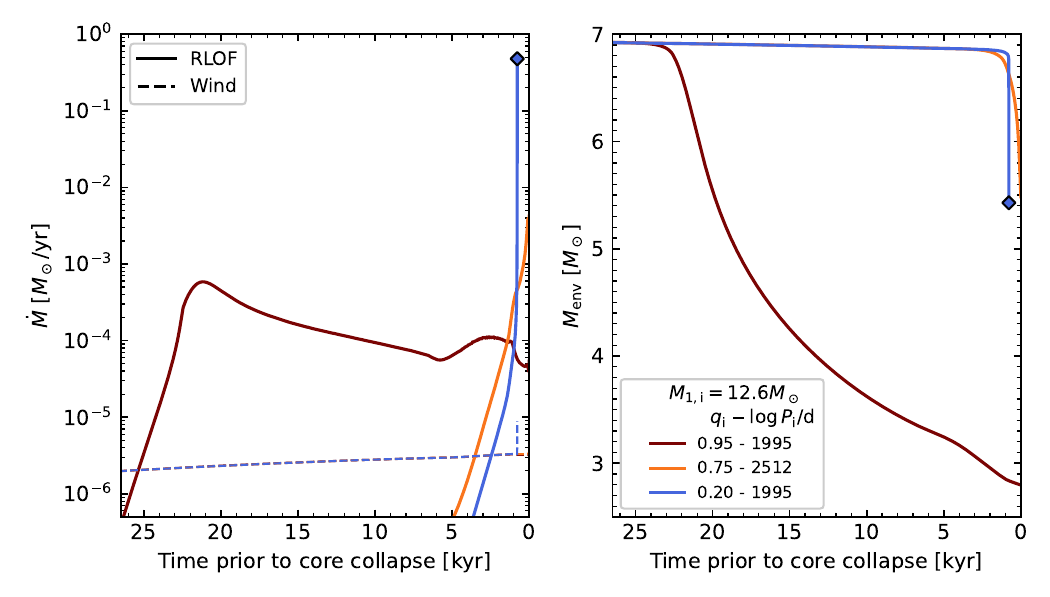}
    \caption{Mass loss rate (left) and envelope mass (right) as a function of time before CC for three exemplary models in the set (highlighted with different colors). In the left panel, both the mass-loss rate by RLOF (solid lines) and winds (dashed lines) are shown. The point where mass transfer turns unstable is marked by a diamond.}
    \label{fig:three_models}
\end{figure}

Here, we will highlight three of the most particular models shown in \cite{Ercolino_widebinary_RSG} (cf. Fig.\,\ref{fig:three_models}), and we will briefly discuss the bulk of the results in Sect.\,\ref{sec:overview}.

\subsection{Model 1}

In this model ($q_\mathrm{i}=0.95$, $P_\mathrm{i}=1995\,\mathrm{d}$) mass transfer begins about $25\,\mathrm{kyr}$ before CC, peaks at a rate of about $10^{-3}\,M_\odot\,\mathrm{yr}^{-1}$ which  then plateaus at roughly $10^{-4}\,M_\odot\,\mathrm{yr}^{-1}$. The star keeps filling its Roche lobe until CC, with roughly the same mass-loss rate, which is about one order of magnitude higher than the wind mass-loss rate. The phase of RLOF removed more than half of its envelope, reducing it to only $2.8M_\odot$ from the initial $6.9M_\odot$. The material that is lost will produce a nearby CSM of almost $4M_\odot$. If this dense material remains close to the binary (e.g. in a circumbinary disk, see Sect.\,\ref{sec:csm}), the SN-explosion would likely resemble a Type IIn SN.

\subsection{Model 2}

This model ($q_\mathrm{i}=0.75$, $P_\mathrm{i}=2512\,\mathrm{d}$) fills its Roche-lobe only in the last $\sim 5\,\mathrm{kyr}$ before CC and achieves significant mass transfer rates soon after. The mass transfer rate here exceeds $10^{-3}M_\odot \, \mathrm{yr}^{-1}$ in the last $1\,\mathrm{kyr}$ before CC, and even peaks at CC, removing a total of $\sim 1.4M_\odot$ of material. This would result in a significantly denser and closer CSM compared to Model 1.

\subsection{Model 3}

This model ($q_\mathrm{i}=0.25$, $P_\mathrm{i}=1995\,\mathrm{d}$) initially has a similar evolution to Model 2, filling its Roche-lobe at about the same time before CC. In here however, the smaller mass of the companion results in significantly higher mass transfer rates up until $\sim 800\,\mathrm{yr}$ before CC, where mass transfer turns unstable, marking the beginning of the phase of common envelope evolution. Given the short timescale between the onset of the common envelope phase and CC, this may imply that the common envelope phase may still be ongoing at the time of CC and, therefore the bulk of the common envelope's mass may still be close to the system by the time of CC. 
At the time of the explosion, the ejecta (which may still contain some leftover H, or be completely H-free) of $\sim4M_\odot$ impacts with $\sim 7M_\odot$ of CSM (i.e., the total mass of the envelope of the RSG) which would convert a significant amount of the ejecta's kinetic energy into radiation - likely producing a super-luminous supernova.

\subsection{Overview of all models and SN features}\label{sec:overview}

These three models show the potential of the binary channel, as different initial binary configurations alone can drive qualitatively different CSM structures as well as progenitor properties. Of the models in the grid in \cite{Ercolino_widebinary_RSG} that undergo Case C mass transfer, only half become unstable, while the remaining models are stable and continue to fill their Roche-lobe until CC  (e.g., Fig.\,\ref{fig:three_models}). The different initial orbital configurations result in different levels of stripping by the time of CC, producing a smooth transition between fully stripped Type Ibc SN-progenitors, partially stripped Type IIb SNe and poorly (or not) stripped Type IIP/L SNe.  

All models undergoing stable Case C RLOF explode while mass transfer is still ongoing, and will likely exhibit a nearby and dense CSM (each of different masses) at the moment of the explosion. In some cases, the mass for the CSM is a significant fraction of the mass of the ejecta of the SN, which would be able to convert a fraction of kinetic energy into radiation. It is therefore reasonable to assume that many of these progenitors may indeed explode as a Type IIn SN. A similar argument can be made for the models undergoing unstable mass transfer, where the duration of the common envelope phase is uncertain, but the contrast between the ejecta and the CSM masses is more significant and may even explain some SLSNe.

The models undergoing Case C RLOF and then exploding as an interacting SN make up about $\sim 5\%$ of all CCSNe coming from the binaries with $M_\mathrm{1,i}=12.6M_\odot$ \cite{Ercolino_widebinary_RSG}. This number is a good order-of-magnitude estimate of the real rate of this phenomenon if we were to include also different masses, and it is within a factor of a few that of the observed fraction of Type IIn SNe reported in \cite{Smartt_rev_2009,Perley2020_ZTF_SNdemographics}. As we will see in Sect.\,\ref{sec:uncertanties}, there are several uncertain physics and assumptions in the calculation that may increase our predicted value closer to that reported in the literature.

\section{Model Uncertainties}\label{sec:uncertanties}
The models presented here adopt a series of assumptions and simplifications in the physics of initially wide binaries, and the structure of RSGs and the CSM. We will discuss some of these points here.

\subsection{Initial Eccentricity}
The models shown in \cite{Ercolino_widebinary_RSG} assume a circularized orbit at ZAMS. Wide binary orbits are not expected to be initially circular due to the lack of processes that can aid in circularizing the orbit before any mass transfer event (as is the case in tighter binaries due to tides). Wide and eccentric binaries systems containing a RSG undergoing episodic mass transfer are observed, like VV-Cep \cite{PollmannBennet2020_VVCep_spec}.   

The effects of eccentricity would be several. Firstly, mass transfer would be periodic (at each periastron passage), and the ejected material may produce distinct shells or even be asymmetrical, which would be then reflected in the SN during interaction. Secondly, mass transfer can be triggered in wider systems than those shown in \cite{Ercolino_widebinary_RSG}, as the periastron distance decreases with increasing eccentricity. This would increase the parameter space of mass-transferring RSG binaries before CC. 

This last point will inevitably also blend the parameter space for earlier mass transfer phases, as adding eccentricity to our models may trigger Case B RLOF instead of Case C, due to the smaller periastron distance.

\subsection{Pulsations}
The variability of all known RSGs \cite{MA23_StellarVariability_GaiaDR3} implies that they exhibit radial pulsations, a phenomenon that is also well studied in the literature through hydrodymic stellar evolution models \cite{Heger1997_Pulsation, Yoon_Cantiello_2010, Moriya_Langer_2015_Pulsations}, where the pulsations are found to grow more significantly with higher $L/M$. This parameter is enhanced in our models (by a factor of two and even more, compared to a single star) thanks to the partial stripping of the RSG during RLOF. 

Pulsations may increase the mass-loss rate in single RSGs, while in binaries the effect would be more dramatic as the change in radius may allow the star to undergo episodic phases of RLOF. The resulting features of the CSM would be qualitatively similar to that of the episodic phases of RLOF during periastron passage in eccentric binaries, except it would be present in circular orbits as well. 

Given the models presented, the pulsations may increase the parameter space for Case C RLOF, as the tightest models not undergoing any RLOF may undergo episodic phases of mass transfer triggered by the pulsations.  

\subsection{Extended Atmospheres}

\begin{figure}
    \centering
    \includegraphics[width=0.8\linewidth]{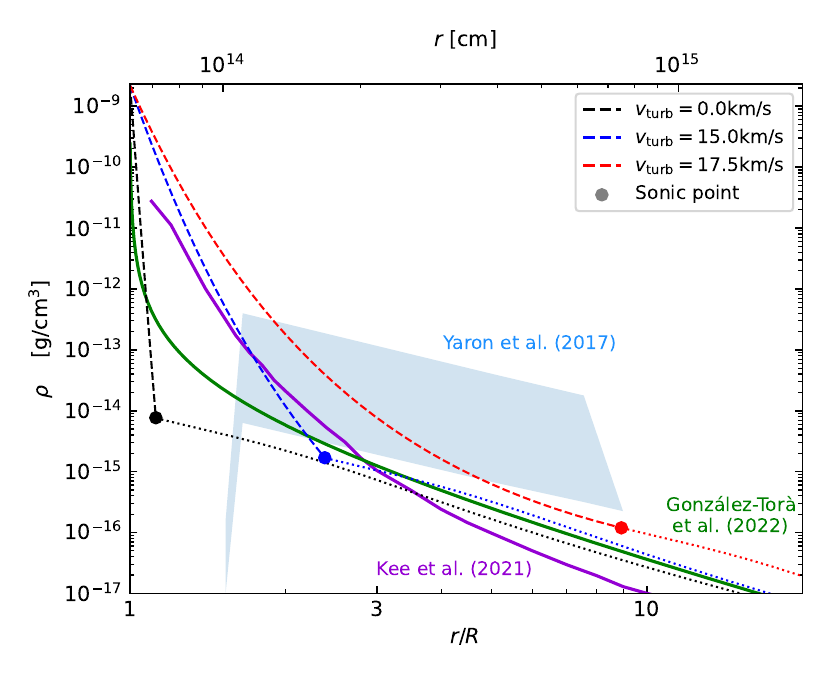}
    \caption{Circumstellar density as a function of the distance from the outer boundary of the RSG single star model taken $1\,\mathrm{yr}$ before CC (see text). The isothermal atmosphere models also include the contribution from turbulent pressure (dashed lines). At the sonic point (bullet scatter), they transition to a $\beta$-wind law (dotted lines). The fiducial turbulent pressure-driven RSG wind model of \cite{Kee_Atmo_Pturb_2021} is shown (solid violet line), as well as the semi-empirical outflowing atmosphere fit to the interferometric observations of  RSG\,HD95687 from \cite{Gonzalez-Tora_RSG} (green solid line). The constraint on the circumstellar density derived from the flash-ionization features in SN\,2013fs is reported (blue area). Figure adapted from \cite{Ercolino_widebinary_RSG}.}
    \label{fig:atmo}
\end{figure}

The stellar models assume there is no matter outside the ``surface'' of the stellar model, defined as where the structure turns optically thin ($\tau<2/3$ \cite{MESA_I}). This is only an approximation, and interferometric observations have suggested the presence of extended, optically-thin, quasi-stationary atmospheres around RSGs \cite{ArroyoTorres_whylargeRSG}.

We therefore ran a hydrodynamical single-star model (of initial mass $12.6M_\odot$, radius $897R_\odot$ with a steady-state wind mass-loss rate of $\sim 3\times 10^{-6}M_\odot\,\mathrm{yr}^{-1}$), which yielded a surface outflow velocity at the surface in the order of $3\,\mathrm{cm}\,\mathrm{s}^{-1}$, which is significantly smaller than the local sound speed ($\sim 5\,\mathrm{km}\,\mathrm{s}^{-1}$), implying the presence of an optically thin but quasi-hydrostatic atmosphere. Assuming an isothermal stratification, and integrating the hydrostatic equation, yields that this atmosphere extends for a region about $\sim 10\%$ the stellar radius before reaching the sonic point. 

However, the deep convective envelope and the turbulence that arises from the convective motion can contribute to the pressure in these layers \cite{Jian_Huang_Pturb_I, Stothers_Turbolence, Grassitelli_Turbolence, Goldberg2022_3Dhydro_turb}. Assuming the contribution is of the form $P_\mathrm{turb}\simeq \frac 1 3 \rho v_\mathrm{turb}^2$, with $v_\mathrm{turb}$ as a constant parameter, the density structure can be again integrated. Using values around $15\,\mathrm{km}\,\mathrm{s}^{-1}$, as proposed in \cite{Kee_Atmo_Pturb_2021}, the sonic point moves as far as a few times the stellar radius. 

These extended atmospheres are relatively dense, and can therefore help trigger mass transfer even if the star is significantly smaller than its Roche-lobe (see also wind-RLOF  \citep{Modamed_Podsiadlowski_WindAccretion_2007, Modamed_WindAccretion_2010}), therefore increasing the parameter space for initially wide mass-transferring systems. If we assume that the hydrostatic atmospheres of RSGs can extend for as much as $\sim 5$ times the modelled radius, the number of interacting supernovae can increase by about $\sim 10-15\%$, bringing this in line with the observed fraction of Type IIn SNe from \cite{Smartt_rev_2009,Perley2020_ZTF_SNdemographics}.

 These extended atmospheres may also help explain a different class of SNe. The densities and extensions derived for these atmospheres are compatible with those inferred in the observations of the flash-ionized SN\,2013fs \cite{Yaron_Flash}. Therefore, if these extended atmospheres are present in RSGs, they may help explain the fact that many Type II SNe, if observed early enough, show flash-ionization features \cite{Bruch2022_CSM_II}.

\subsection{CSM geometry}\label{sec:csm}
{  Typically, works discussing SN-CSM interaction assume that the CSM is distributed spherically around the progenitor star \cite{Chatzopoulos2013_analiticalmodel, Villar17_lightcurves,MOSFIT}. This is sensible for a CSM produced by a single star in the context of a sudden phase of mass loss, but it is worth noting that the multidimensional nature of RSG outflows may induce significant deviations from spherical symmetry. If these outflows are the result of binary interaction, the CSM need not be spherical, depending on the ejection mechanism.} Some works suggest, in the case of a low-mass giant donor, that the mass-outflows from inefficient mass-transfer tend to remain confined towards the orbital plane (e.g., \cite{TheunsJorissen1993_WindAccretionBinaries, MastrodemosMorris1998_BipolarNebulaeBinaries, Walder2008_RSOph_Nova_accretion, BoothMohamedPods_2016_innertorque_CSM_RSoph_Ia}). Here, since the mass loss rates are much higher, the outflow velocities are expected to be slower and the material may not escape the system and instead form a circumbinary disk. Such configuration usually sees a cavity until  distances of about $2-3$ times the binary separation. Such disks in our models would therefore be present for distances above $\sim 3\times 10^{14}\,\mathrm{cm}$, which is within the range of distances expected for the CSM in Type IIn SNe. {  The presence of such a disk-like CSM would also affect the evolution of the angular momentum of the inner binary by tides and torques, and possibly even pump eccentricity \cite{Valli24_CBD_binaryevolution}.}

The asphericity of the CSM will also include additional features on the observed SN, both in its light-curve and spectra, also in terms of the viewing-angle of the disk \cite{Vlasis16_asymmetricmodels_IIn, Suzuki2019_HrichSN2DDiskCSM} {  and polarization features \citep{Mauerhan2014_aymmCSM_spectropol}}. This would further contribute to the diversity of the exploding SN from these systems.

\section{Conclusion}
In this work, we summarized the key results from \cite{Ercolino_widebinary_RSG}, which focused on the effects of mass transfer in wide binaries where the donor, and SN-progenitor, is a RSG star. 

The work finds that mass transfer from a RSG can be stable, and produces a smooth transition between different SN types depending on the initial orbital configuration. If Case C RLOF is stable, the progenitor star continues to fill its Roche lobe until the moment of explosion, likely resulting in a dense and close CSM that will affect the observed SN features. If Case C RLOF turns unstable, the resulting SN will likely be very bright, due to the presence of a very massive CSM.

At the same time, \cite{Ercolino_widebinary_RSG} also shows that several considerations on the assumptions in the calculations of these models (namely the initial eccentricity of the binary as well as the pulsation of the envelope and extension of the atmosphere of RSGs) may increase the number of systems undergoing Case C RLOF, and therefore result in even more interacting SN than the fiducial value of $5\%$ that was derived.  

Finally, the CSM that is produced in each system is different, from having different masses to having been produced with different mechanisms and in different timescales. The pulsations in RSGs and the initial eccentricity of the binaries they would be found in may also contribute to the diversity of the CSM by adding episodic phases of mass-loss. In the case in which the CSM is found in a disk, as it would likely be expected in such binaries, viewing-angle dependent effects may also play a role in the observation of these SNe.



\begin{furtherinformation}

\begin{orcids}

\orcid{0000-0002-2807-5253}{Andrea}{Ercolino}
\end{orcids}


\begin{conflictsofinterest}
The author declares that there is no conflict of interest.
\end{conflictsofinterest}

\end{furtherinformation}



%

\bibliographystyle{bullsrsl-numen}

\bibliography{A_ERCOLINO_references}

\begin{thebibliography}{10}
\providecommand{\url}[1]{#1}
\providecommand{\urlprefix}{URL }

\bibitem{ASASN}
{Shappee}, B., {Prieto}, J., {Stanek}, K.~Z., {Kochanek}, C.~S., {Holoien}, T., {Jencson}, J., {Basu}, U., {Beacom}, J.~F., {Szczygiel}, D., {Pojmanski}, G., {Brimacombe}, J., {Dubberley}, M., {Elphick}, M., {Foale}, S., {Hawkins}, E., {Mullins}, D., {Rosing}, W., {Ross}, R. and {Walker}, Z. (2014) {All Sky Automated Survey for SuperNovae (ASAS-SN or ``Assassin'')}.
\newblock In American Astronomical Society Meeting Abstracts \#223, vol. 223 of \emph{American Astronomical Society Meeting Abstracts}, p. 236.03.

\bibitem{ATLAS}
{Tonry}, J.~L., {Denneau}, L., {Heinze}, A.~N., {Stalder}, B., {Smith}, K.~W., {Smartt}, S.~J., {Stubbs}, C.~W., {Weiland}, H.~J. and {Rest}, A. (2018) {ATLAS: A High-cadence All-sky Survey System}.
\newblock Publications of the ASP, 130(988), 064505.
\newblock \url{https://doi.org/10.1088/1538-3873/aabadf}.

\bibitem{PanSTARRS}
{Kaiser}, N., {Burgett}, W., {Chambers}, K., {Denneau}, L., {Heasley}, J., {Jedicke}, R., {Magnier}, E., {Morgan}, J., {Onaka}, P. and {Tonry}, J. (2010) {The Pan-STARRS wide-field optical/NIR imaging survey}.
\newblock In Ground-based and Airborne Telescopes III, edited by {Stepp}, L.~M., {Gilmozzi}, R. and {Hall}, H.~J., vol. 7733 of \emph{Society of Photo-Optical Instrumentation Engineers (SPIE) Conference Series}, p. 77330E.
\newblock \url{https://doi.org/10.1117/12.859188}.

\bibitem{PTF}
{Law}, N.~M., {Kulkarni}, S.~R., {Dekany}, R.~G., {Ofek}, E.~O., {Quimby}, R.~M., {Nugent}, P.~E., {Surace}, J., {Grillmair}, C.~C., {Bloom}, J.~S., {Kasliwal}, M.~M., {Bildsten}, L., {Brown}, T., {Cenko}, S.~B., {Ciardi}, D., {Croner}, E., {Djorgovski}, S.~G., {van Eyken}, J., {Filippenko}, A.~V., {Fox}, D.~B., {Gal-Yam}, A., {Hale}, D., {Hamam}, N., {Helou}, G., {Henning}, J., {Howell}, D.~A., {Jacobsen}, J., {Laher}, R., {Mattingly}, S., {McKenna}, D., {Pickles}, A., {Poznanski}, D., {Rahmer}, G., {Rau}, A., {Rosing}, W., {Shara}, M., {Smith}, R., {Starr}, D., {Sullivan}, M., {Velur}, V., {Walters}, R. and {Zolkower}, J. (2009) {The Palomar Transient Factory: System Overview, Performance, and First Results}.
\newblock Publications of the ASP, 121(886), 1395.
\newblock \url{https://doi.org/10.1086/648598}.

\bibitem{ZTF}
{Bellm}, E.~C., {Kulkarni}, S.~R., {Graham}, M.~J., {Dekany}, R., {Smith}, R.~M., {Riddle}, R., {Masci}, F.~J., {Helou}, G., {Prince}, T.~A., {Adams}, S.~M., {Barbarino}, C., {Barlow}, T., {Bauer}, J., {Beck}, R., {Belicki}, J., {Biswas}, R., {Blagorodnova}, N., {Bodewits}, D., {Bolin}, B., {Brinnel}, V., {Brooke}, T., {Bue}, B., {Bulla}, M., {Burruss}, R., {Cenko}, S.~B., {Chang}, C.-K., {Connolly}, A., {Coughlin}, M., {Cromer}, J., {Cunningham}, V., {De}, K., {Delacroix}, A., {Desai}, V., {Duev}, D.~A., {Eadie}, G., {Farnham}, T.~L., {Feeney}, M., {Feindt}, U., {Flynn}, D., {Franckowiak}, A., {Frederick}, S., {Fremling}, C., {Gal-Yam}, A., {Gezari}, S., {Giomi}, M., {Goldstein}, D.~A., {Golkhou}, V.~Z., {Goobar}, A., {Groom}, S., {Hacopians}, E., {Hale}, D., {Henning}, J., {Ho}, A. Y.~Q., {Hover}, D., {Howell}, J., {Hung}, T., {Huppenkothen}, D., {Imel}, D., {Ip}, W.-H., {Ivezi{\'c}}, {\v{Z}}., {Jackson}, E., {Jones}, L., {Juric}, M., {Kasliwal}, M.~M., {Kaspi}, S., {Kaye}, S., {Kelley}, M. S.~P.,
  {Kowalski}, M., {Kramer}, E., {Kupfer}, T., {Landry}, W., {Laher}, R.~R., {Lee}, C.-D., {Lin}, H.~W., {Lin}, Z.-Y., {Lunnan}, R., {Giomi}, M., {Mahabal}, A., {Mao}, P., {Miller}, A.~A., {Monkewitz}, S., {Murphy}, P., {Ngeow}, C.-C., {Nordin}, J., {Nugent}, P., {Ofek}, E., {Patterson}, M.~T., {Penprase}, B., {Porter}, M., {Rauch}, L., {Rebbapragada}, U., {Reiley}, D., {Rigault}, M., {Rodriguez}, H., {van Roestel}, J., {Rusholme}, B., {van Santen}, J., {Schulze}, S., {Shupe}, D.~L., {Singer}, L.~P., {Soumagnac}, M.~T., {Stein}, R., {Surace}, J., {Sollerman}, J., {Szkody}, P., {Taddia}, F., {Terek}, S., {Van Sistine}, A., {van Velzen}, S., {Vestrand}, W.~T., {Walters}, R., {Ward}, C., {Ye}, Q.-Z., {Yu}, P.-C., {Yan}, L. and {Zolkower}, J. (2019) {The Zwicky Transient Facility: System Overview, Performance, and First Results}.
\newblock Publications of the ASP, 131(995), 018002.
\newblock \url{https://doi.org/10.1088/1538-3873/aaecbe}.

\bibitem{Fransson_2010jl}
{Fransson}, C., {Ergon}, M., {Challis}, P.~J., {Chevalier}, R.~A., {France}, K., {Kirshner}, R.~P., {Marion}, G.~H., {Milisavljevic}, D., {Smith}, N., {Bufano}, F., {Friedman}, A.~S., {Kangas}, T., {Larsson}, J., {Mattila}, S., {Benetti}, S., {Chornock}, R., {Czekala}, I., {Soderberg}, A. and {Sollerman}, J. (2014) {High-density Circumstellar Interaction in the Luminous Type IIn SN 2010jl: The First 1100 Days}.
\newblock Astrophysical Journal, 797(2), 118.
\newblock \url{https://doi.org/10.1088/0004-637X/797/2/118}.

\bibitem{Smith2008_2006gy}
{Smith}, N., {Foley}, R.~J., {Bloom}, J.~S., {Li}, W., {Filippenko}, A.~V., {Gavazzi}, R., {Ghez}, A., {Konopacky}, Q., {Malkan}, M.~A., {Marshall}, P.~J., {Pooley}, D., {Treu}, T. and {Woo}, J.-H. (2008) {Late-Time Observations of SN 2006gy: Still Going Strong}.
\newblock The Astrophysical Journal, 686(1), 485--491.
\newblock \url{https://doi.org/10.1086/590141}.

\bibitem{Yaron_Flash}
{Yaron}, O., {Perley}, D.~A., {Gal-Yam}, A., {Groh}, J.~H., {Horesh}, A., {Ofek}, E.~O., {Kulkarni}, S.~R., {Sollerman}, J., {Fransson}, C., {Rubin}, A., {Szabo}, P., {Sapir}, N., {Taddia}, F., {Cenko}, S.~B., {Valenti}, S., {Arcavi}, I., {Howell}, D.~A., {Kasliwal}, M.~M., {Vreeswijk}, P.~M., {Khazov}, D., {Fox}, O.~D., {Cao}, Y., {Gnat}, O., {Kelly}, P.~L., {Nugent}, P.~E., {Filippenko}, A.~V., {Laher}, R.~R., {Wozniak}, P.~R., {Lee}, W.~H., {Rebbapragada}, U.~D., {Maguire}, K., {Sullivan}, M. and {Soumagnac}, M.~T. (2017) {Confined dense circumstellar material surrounding a regular type II supernova}.
\newblock Nature Physics, 13(5), 510--517.
\newblock \url{https://doi.org/10.1038/nphys4025}.

\bibitem{Margutti2017_2014C}
{Margutti}, R., {Kamble}, A., {Milisavljevic}, D., {Zapartas}, E., {de Mink}, S.~E., {Drout}, M., {Chornock}, R., {Risaliti}, G., {Zauderer}, B.~A., {Bietenholz}, M., {Cantiello}, M., {Chakraborti}, S., {Chomiuk}, L., {Fong}, W., {Grefenstette}, B., {Guidorzi}, C., {Kirshner}, R., {Parrent}, J.~T., {Patnaude}, D., {Soderberg}, A.~M., {Gehrels}, N.~C. and {Harrison}, F. (2017) {Ejection of the Massive Hydrogen-rich Envelope Timed with the Collapse of the Stripped SN 2014C}.
\newblock Astrophysical Journal, 835(2), 140.
\newblock \url{https://doi.org/10.3847/1538-4357/835/2/140}.

\bibitem{Leonard_1998S}
{Leonard}, D.~C., {Filippenko}, A.~V., {Barth}, A.~J. and {Matheson}, T. (2000) {Evidence for Asphericity in the Type IIN Supernova SN 1998S}.
\newblock Astrophysical Journal, 536(1), 239--254.
\newblock \url{https://doi.org/10.1086/308910}.

\bibitem{SmithArnett2014_Hydroinstab_Turb_preSN}
{Smith}, N. and {Arnett}, W.~D. (2014) {Preparing for an Explosion: Hydrodynamic Instabilities and Turbulence in Presupernovae}.
\newblock Astrophysical Journal, 785(2), 82.
\newblock \url{https://doi.org/10.1088/0004-637X/785/2/82}.

\bibitem{Quataert_Shiode_wavedriven_winds}
{Quataert}, E. and {Shiode}, J. (2012) {Wave-driven mass loss in the last year of stellar evolution: setting the stage for the most luminous core-collapse supernovae}.
\newblock Monthly Notices of the RAS, 423(1), L92--L96.
\newblock \url{https://doi.org/10.1111/j.1745-3933.2012.01264.x}.

\bibitem{Woosley_Heger_2015_SiFlash}
{Woosley}, S.~E. and {Heger}, A. (2015) {The Remarkable Deaths of 9-11 Solar Mass Stars}.
\newblock Astrophysical Journal, 810(1), 34.
\newblock \url{https://doi.org/10.1088/0004-637X/810/1/34}.

\bibitem{Fuller17_waveheating_RSG}
{Fuller}, J. (2017) {Pre-supernova outbursts via wave heating in massive stars - I. Red supergiants}.
\newblock Monthly Notices of the RAS, 470(2), 1642--1656.
\newblock \url{https://doi.org/10.1093/mnras/stx1314}.

\bibitem{Sana_massive_stars_binaries}
{Sana}, H., {de Mink}, S.~E., {de Koter}, A., {Langer}, N., {Evans}, C.~J., {Gieles}, M., {Gosset}, E., {Izzard}, R.~G., {Le Bouquin}, J.~B. and {Schneider}, F.~R.~N. (2012) {Binary Interaction Dominates the Evolution of Massive Stars}.
\newblock Science, 337(6093), 444.
\newblock \url{https://doi.org/10.1126/science.1223344}.

\bibitem{Ercolino_widebinary_RSG}
{Ercolino}, A., {Jin}, H., {Langer}, N. and {Dessart}, L. (2024) {Interacting supernovae from wide massive binary systems}.
\newblock Astronomy \& Astrophysics, 685, A58.
\newblock \url{https://doi.org/10.1051/0004-6361/202347646}.

\bibitem{MESA_I}
{Paxton}, B., {Bildsten}, L., {Dotter}, A., {Herwig}, F., {Lesaffre}, P. and {Timmes}, F. (2011) {Modules for Experiments in Stellar Astrophysics (MESA)}.
\newblock Astrophysical Journal, Supplement, 192(1), 3.
\newblock \url{https://doi.org/10.1088/0067-0049/192/1/3}.

\bibitem{MESA_II}
{Paxton}, B., {Cantiello}, M., {Arras}, P., {Bildsten}, L., {Brown}, E.~F., {Dotter}, A., {Mankovich}, C., {Montgomery}, M.~H., {Stello}, D., {Timmes}, F.~X. and {Townsend}, R. (2013) {Modules for Experiments in Stellar Astrophysics (MESA): Planets, Oscillations, Rotation, and Massive Stars}.
\newblock Astrophysical Journal, Supplement, 208(1), 4.
\newblock \url{https://doi.org/10.1088/0067-0049/208/1/4}.

\bibitem{MESA_III}
{Paxton}, B., {Marchant}, P., {Schwab}, J., {Bauer}, E.~B., {Bildsten}, L., {Cantiello}, M., {Dessart}, L., {Farmer}, R., {Hu}, H., {Langer}, N., {Townsend}, R.~H.~D., {Townsley}, D.~M. and {Timmes}, F.~X. (2015) {Modules for Experiments in Stellar Astrophysics (MESA): Binaries, Pulsations, and Explosions}.
\newblock Astrophysical Journal, Supplement, 220(1), 15.
\newblock \url{https://doi.org/10.1088/0067-0049/220/1/15}.

\bibitem{MESA_IV}
{Paxton}, B., {Schwab}, J., {Bauer}, E.~B., {Bildsten}, L., {Blinnikov}, S., {Duffell}, P., {Farmer}, R., {Goldberg}, J.~A., {Marchant}, P., {Sorokina}, E., {Thoul}, A., {Townsend}, R. H.~D. and {Timmes}, F.~X. (2018) {Modules for Experiments in Stellar Astrophysics (MESA): Convective Boundaries, Element Diffusion, and Massive Star Explosions}.
\newblock Astrophysical Journal, Supplement, 234(2), 34.
\newblock \url{https://doi.org/10.3847/1538-4365/aaa5a8}.

\bibitem{MESA_V}
{Paxton}, B., {Smolec}, R., {Schwab}, J., {Gautschy}, A., {Bildsten}, L., {Cantiello}, M., {Dotter}, A., {Farmer}, R., {Goldberg}, J.~A., {Jermyn}, A.~S., {Kanbur}, S.~M., {Marchant}, P., {Thoul}, A., {Townsend}, R. H.~D., {Wolf}, W.~M., {Zhang}, M. and {Timmes}, F.~X. (2019) {Modules for Experiments in Stellar Astrophysics (MESA): Pulsating Variable Stars, Rotation, Convective Boundaries, and Energy Conservation}.
\newblock Astrophysical Journal, Supplement, 243(1), 10.
\newblock \url{https://doi.org/10.3847/1538-4365/ab2241}.

\bibitem{Smartt_rev_2009}
{Smartt}, S.~J. (2009) {Progenitors of Core-Collapse Supernovae}.
\newblock Annual Review of Astron and Astrophysis, 47(1), 63--106.
\newblock \url{https://doi.org/10.1146/annurev-astro-082708-101737}.

\bibitem{Perley2020_ZTF_SNdemographics}
{Perley}, D.~A., {Fremling}, C., {Sollerman}, J., {Miller}, A.~A., {Dahiwale}, A.~S., {Sharma}, Y., {Bellm}, E.~C., {Biswas}, R., {Brink}, T.~G., {Bruch}, R.~J., {De}, K., {Dekany}, R., {Drake}, A.~J., {Duev}, D.~A., {Filippenko}, A.~V., {Gal-Yam}, A., {Goobar}, A., {Graham}, M.~J., {Graham}, M.~L., {Ho}, A. Y.~Q., {Irani}, I., {Kasliwal}, M.~M., {Kim}, Y.-L., {Kulkarni}, S.~R., {Mahabal}, A., {Masci}, F.~J., {Modak}, S., {Neill}, J.~D., {Nordin}, J., {Riddle}, R.~L., {Soumagnac}, M.~T., {Strotjohann}, N.~L., {Schulze}, S., {Taggart}, K., {Tzanidakis}, A., {Walters}, R.~S. and {Yan}, L. (2020) {The Zwicky Transient Facility Bright Transient Survey. II. A Public Statistical Sample for Exploring Supernova Demographics}.
\newblock Astrophysical Journal, 904(1), 35.
\newblock \url{https://doi.org/10.3847/1538-4357/abbd98}.

\bibitem{PollmannBennet2020_VVCep_spec}
{Pollmann}, E. and {Bennett}, P. (2020) {Spectroscopic Monitoring of the 2017-2019 Eclipse of VV Cephei}.
\newblock Journal of the American Association of Variable Star Observers, 48(2), 118.

\bibitem{MA23_StellarVariability_GaiaDR3}
{Ma{\'\i}z Apell{\'a}niz}, J., {Holgado}, G., {Pantaleoni Gonz{\'a}lez}, M. and {Caballero}, J.~A. (2023) {Stellar variability in Gaia DR3. I. Three-band photometric dispersions for 145 million sources}.
\newblock arXiv e-prints, arXiv:2304.14249.
\newblock \url{https://doi.org/10.48550/arXiv.2304.14249}.

\bibitem{Heger1997_Pulsation}
{Heger}, A., {Jeannin}, L., {Langer}, N. and {Baraffe}, I. (1997) {Pulsations in red supergiants with high L/M ratio. Implications for the stellar and circumstellar structure of supernova progenitors}.
\newblock Astronomy \& Astrophysics, 327, 224--230.
\newblock \url{https://doi.org/10.48550/arXiv.astro-ph/9705097}.

\bibitem{Yoon_Cantiello_2010}
{Yoon}, S.-C. and {Cantiello}, M. (2010) {Evolution of Massive Stars with Pulsation-driven Superwinds During the Red Supergiant Phase}.
\newblock Astrophysical Journal Letters, 717(1), L62--L65.
\newblock \url{https://doi.org/10.1088/2041-8205/717/1/L62}.

\bibitem{Moriya_Langer_2015_Pulsations}
{Moriya}, T.~J. and {Langer}, N. (2015) {Pulsations of red supergiant pair-instability supernova progenitors leading to extreme mass loss}.
\newblock Astronomy \& Astrophysics, 573, A18.
\newblock \url{https://doi.org/10.1051/0004-6361/201424957}.

\bibitem{Kee_Atmo_Pturb_2021}
{Kee}, N.~D., {Sundqvist}, J.~O., {Decin}, L., {de Koter}, A. and {Sana}, H. (2021) {Analytic, dust-independent mass-loss rates for red supergiant winds initiated by turbulent pressure}.
\newblock Astronomy \& Astrophysics, 646, A180.
\newblock \url{https://doi.org/10.1051/0004-6361/202039224}.

\bibitem{Gonzalez-Tora_RSG}
{Gonz{\'a}lez-Tor{\`a}}, G., {Wittkowski}, M., {Davies}, B., {Plez}, B. and {Kravchenko}, K. (2023) {The effect of winds on atmospheric layers of red supergiants. I. Modelling for interferometric observations}.
\newblock Astronomy \& Astrophysics, 669, A76.
\newblock \url{https://doi.org/10.1051/0004-6361/202244503}.

\bibitem{ArroyoTorres_whylargeRSG}
{Arroyo-Torres}, B., {Wittkowski}, M., {Chiavassa}, A., {Scholz}, M., {Freytag}, B., {Marcaide}, J.~M., {Hauschildt}, P.~H., {Wood}, P.~R. and {Abellan}, F.~J. (2015) {What causes the large extensions of red supergiant atmospheres?. Comparisons of interferometric observations with 1D hydrostatic, 3D convection, and 1D pulsating model atmospheres}.
\newblock Astronomy \& Astrophysics, 575, A50.
\newblock \url{https://doi.org/10.1051/0004-6361/201425212}.

\bibitem{Jian_Huang_Pturb_I}
{Jiang}, S.~Y. and {Huang}, R.~Q. (1997) {The effect of turbulent pressure on the red giants and AGB stars. I. On the internal structure and evolution.}
\newblock Astronomy \& Astrophysics, 317, 114--120.

\bibitem{Stothers_Turbolence}
{Stothers}, R.~B. (2003) {Turbulent Pressure in the Envelopes of Yellow Hypergiants and Luminous Blue Variables}.
\newblock Astrophysical Journal, 589(2), 960--967.
\newblock \url{https://doi.org/10.1086/374713}.

\bibitem{Grassitelli_Turbolence}
{Grassitelli}, L., {Fossati}, L., {Langer}, N., {Sim{\'o}n-D{\'\i}az}, S., {Castro}, N. and {Sanyal}, D. (2016) {Metallicity dependence of turbulent pressure and macroturbulence in stellar envelopes}.
\newblock Astronomy \& Astrophysics, 593, A14.
\newblock \url{https://doi.org/10.1051/0004-6361/201628912}.

\bibitem{Goldberg2022_3Dhydro_turb}
{Goldberg}, J.~A., {Jiang}, Y.-F. and {Bildsten}, L. (2022) {Numerical Simulations of Convective Three-dimensional Red Supergiant Envelopes}.
\newblock Astrophysical Journal, 929(2), 156.
\newblock \url{https://doi.org/10.3847/1538-4357/ac5ab3}.

\bibitem{Modamed_Podsiadlowski_WindAccretion_2007}
{Mohamed}, S. and {Podsiadlowski}, P. (2007) {Wind Roche-Lobe Overflow: a New Mass-Transfer Mode for Wide Binaries}.
\newblock In 15th European Workshop on White Dwarfs, edited by {Napiwotzki}, R. and {Burleigh}, M.~R., vol. 372 of \emph{Astronomical Society of the Pacific Conference Series}, p. 397.

\bibitem{Modamed_WindAccretion_2010}
{Mohamed}, S. and {Podsiadlowski}, P. (2010) {Understanding Mass Transfer in Wind-Interacting Binaries: SPH Models of ``Wind Roche-lobe Overflow''}.
\newblock In International Conference on Binaries: in celebration of Ron Webbink's 65th Birthday, edited by {Kalogera}, V. and {van der Sluys}, M., vol. 1314 of \emph{American Institute of Physics Conference Series}, pp. 51--52.
\newblock \url{https://doi.org/10.1063/1.3536409}.

\bibitem{Bruch2022_CSM_II}
{Bruch}, R.~J., {Gal-Yam}, A., {Yaron}, O., {Chen}, P., {Strotjohann}, N.~L., {Irani}, I., {Zimmerman}, E., {Schulze}, S., {Yang}, Y., {Kim}, Y.-L., {Bulla}, M., {Sollerman}, J., {Rigault}, M., {Ofek}, E., {Soumagnac}, M., {Masci}, F.~J., {Fremling}, C., {Perley}, D., {Nordin}, J., {Cenko}, S.~B., {Ho}, A. Y.~Q., {Adams}, S., {Adreoni}, I., {Bellm}, E.~C., {Blagorodnova}, N., {Burdge}, K., {De}, K., {Dekany}, R.~G., {Dhawan}, S., {Drake}, A.~J., {Duev}, D.~A., {Graham}, M., {Graham}, M.~L., {Jencson}, J., {Karamehmetoglu}, E., {Kasliwal}, M.~M., {Kulkarni}, S., {Miller}, A.~A., {Neill}, J.~D., {Prince}, T.~A., {Riddle}, R., {Rusholme}, B., {Sharma}, Y., {Smith}, R., {Sravan}, N., {Taggart}, K., {Walters}, R. and {Yan}, L. (2023) {The Prevalence and Influence of Circumstellar Material around Hydrogen-rich Supernova Progenitors}.
\newblock Astrophysical Journal, 952(2), 119.
\newblock \url{https://doi.org/10.3847/1538-4357/acd8be}.

\bibitem{Chatzopoulos2013_analiticalmodel}
{Chatzopoulos}, E., {Wheeler}, J.~C., {Vinko}, J., {Horvath}, Z.~L. and {Nagy}, A. (2013) {Analytical Light Curve Models of Superluminous Supernovae: {\ensuremath{\chi}}$^{2}$-minimization of Parameter Fits}.
\newblock Astrophysical Journal, 773(1), 76.
\newblock \url{https://doi.org/10.1088/0004-637X/773/1/76}.

\bibitem{Villar17_lightcurves}
{Villar}, V.~A., {Berger}, E., {Metzger}, B.~D. and {Guillochon}, J. (2017) {Theoretical Models of Optical Transients. I. A Broad Exploration of the Duration-Luminosity Phase Space}.
\newblock Astrophysical Journal, 849(1), 70.
\newblock \url{https://doi.org/10.3847/1538-4357/aa8fcb}.

\bibitem{MOSFIT}
{Guillochon}, J., {Nicholl}, M., {Villar}, V.~A., {Mockler}, B., {Narayan}, G., {Mandel}, K.~S., {Berger}, E. and {Williams}, P. K.~G. (2018) {MOSFiT: Modular Open Source Fitter for Transients}.
\newblock Astrophysical Journal, Supplement, 236(1), 6.
\newblock \url{https://doi.org/10.3847/1538-4365/aab761}.

\bibitem{TheunsJorissen1993_WindAccretionBinaries}
{Theuns}, T. and {Jorissen}, A. (1993) {Wind accretion in binary stars - I. Intricacies of the flow structure.}
\newblock Monthly Notices of the RAS, 265, 946--967.
\newblock \url{https://doi.org/10.1093/mnras/265.4.946}.

\bibitem{MastrodemosMorris1998_BipolarNebulaeBinaries}
{Mastrodemos}, N. and {Morris}, M. (1998) {Bipolar Preplanetary Nebulae: Hydrodynamics of Dusty Winds in Binary Systems. I. Formation of Accretion Disks}.
\newblock Astrophysical Journal, 497(1), 303--329.
\newblock \url{https://doi.org/10.1086/305465}.

\bibitem{Walder2008_RSOph_Nova_accretion}
{Walder}, R., {Folini}, D. and {Shore}, S.~N. (2008) {3D simulations of RS Ophiuchi: from accretion to nova blast}.
\newblock Astronomy \& Astrophysics, 484(1), L9--L12.
\newblock \url{https://doi.org/10.1051/0004-6361:200809703}.

\bibitem{BoothMohamedPods_2016_innertorque_CSM_RSoph_Ia}
{Booth}, R.~A., {Mohamed}, S. and {Podsiadlowski}, P. (2016) {Modelling the circumstellar medium in RS Ophiuchi and its link to Type Ia supernovae}.
\newblock Monthly Notices of the RAS, 457(1), 822--835.
\newblock \url{https://doi.org/10.1093/mnras/stw001}.

\bibitem{Valli24_CBD_binaryevolution}
{Valli}, R., {Tiede}, C., {Vigna-G{\'o}mez}, A., {Cuadra}, J., {Siwek}, M., {Ma}, J.-Z., {D'Orazio}, D.~J., {Zrake}, J. and {de Mink}, S.~E. (2024) {Long-term Evolution of Binary Orbits Induced by Circumbinary Disks}.
\newblock arXiv e-prints, arXiv:2401.17355.
\newblock \url{https://doi.org/10.48550/arXiv.2401.17355}.

\bibitem{Vlasis16_asymmetricmodels_IIn}
{Vlasis}, A., {Dessart}, L. and {Audit}, E. (2016) {Two-dimensional radiation hydrodynamics simulations of superluminous interacting supernovae of Type IIn}.
\newblock Monthly Notices of the RAS, 458(2), 1253--1266.
\newblock \url{https://doi.org/10.1093/mnras/stw410}.

\bibitem{Suzuki2019_HrichSN2DDiskCSM}
{Suzuki}, A., {Moriya}, T.~J. and {Takiwaki}, T. (2019) {Supernova Ejecta Interacting with a Circumstellar Disk. I. Two-dimensional Radiation-hydrodynamic Simulations}.
\newblock Astrophysical Journal, 887(2), 249.
\newblock \url{https://doi.org/10.3847/1538-4357/ab5a83}.

\bibitem{Mauerhan2014_aymmCSM_spectropol}
{Mauerhan}, J., {Williams}, G.~G., {Smith}, N., {Smith}, P.~S., {Filippenko}, A.~V., {Hoffman}, J.~L., {Milne}, P., {Leonard}, D.~C., {Clubb}, K.~I., {Fox}, O.~D. and {Kelly}, P.~L. (2014) {Multi-epoch spectropolarimetry of SN 2009ip: direct evidence for aspherical circumstellar material}.
\newblock Monthly Notices of the RAS, 442(2), 1166--1180.
\newblock \url{https://doi.org/10.1093/mnras/stu730}.

\end{thebibliography}

\end{document}